\documentclass{article}
\usepackage{spconf,amsmath,graphicx}
\usepackage{array}
\usepackage{url}
\newcommand{\B}{\fontseries{b}\selectfont}
\title{UNCERTAINTY-BASED ENSEMBLE LEARNING FOR SPEECH CLASSIFICATION}
\twoauthors
 {Bagus Tris Atmaja \sthanks{Based on project JPNP20006 funded by NEDO Japan.}}
 {AIST\\
 Tsukuba\\
 Japan}
 {Felix Burkhardt}
 {audEERING GmbH\\
 Gilching\\
 Germany}
\begin{document}
%\ninept
%
\maketitle
\begin{abstract}
Speech classification has attracted increasing attention due to its wide applications, particularly in classifying physical and mental states. However, these tasks are challenging due to the high variability in speech signals. Ensemble learning has shown promising results when multiple classifiers are combined to improve performance. With recent advancements in hardware development, combining several models is not a limitation in deep learning research and applications. In this paper, we propose an uncertainty-based ensemble learning approach for speech classification. Specifically, we train a set of base features on the same classifier and quantify the uncertainty of their predictions. The predictions are combined using variants of uncertainty calculation to produce the final prediction. The visualization of the effect of uncertainty and its ensemble learning results show potential improvements in speech classification tasks. The proposed method outperforms single models and conventional ensemble learning methods in terms of unweighted accuracy or weighted accuracy.

\end{abstract}
\section{Introduction}
\label{sec:intro}

Speech classification is a task that classifies speech signals into a set of predefined classes. The task is challenging due to the variability of speech signals caused by various factors such as speaker, environment, and channel. Ensemble learning has been widely used in speech classification to address these challenges. Ensemble learning is a machine learning technique that combines multiple models to improve the system's performance. This technique is known to be effective in speech and audio classifications \cite{Ristea2021,Shih2017,Atmaja2023a}.

On the other hand, every scenario in life deals with uncertainties, including speech classification. Quantifying uncertainty becomes inseparable in machine learning and deep learning techniques \cite{Abdar2021}. Since the prediction of deep learning models includes uncertainty, the performance of the model measured by the accuracy might be affected by the uncertainty score. Given several models for the same task, the uncertainty of each model is different. Suppose that different models output different uncertainty values in a task. Ensembling several models is a potential solution to reduce uncertainty by choosing the lowest uncertainty value across models or weighting the prediction with uncertainty values.

We propose uncertainty-based ensemble learning methods for speech classification. The main contribution is the evaluation of four variants of uncertainties for ensemble learning. The first proposed method uses the lowest uncertainty estimates to select the most confident models for the final decision. We then add a threshold in the second method; higher uncertainties use the mean ensemble. In the third and fourth methods, we use inverse and complement of uncertainty to weigh model probabilities. We evaluate the proposed methods on a speech classification task and compare them with mean and max ensemble learning methods. We count the number of improvements (in terms of weighted and unweighted accuracies) for each method to justify which method is better than other methods.

% In \cite{Atmaja2020} the authors proposed very sophisticated technique.

\section{Methods}
\label{sec:method}
\subsection{Uncertainty Quantification}
Uncertainty quantification (UQ) is an important indicator for trustworthy deep learning. Previous studies have shown that uncertainty quantification can impact the performance of speech emotion recognition models \cite{Schrufer2024}. Guided by that study, we expand the application of uncertainty quantification to ensemble learning for speech classification.

We approach uncertainty quantification by using the entropy of the probabilities (logits). The entropy is calculated as follows: $H(i) = - \sum_{i=1}^N p_i\log(p_i)$, where $p_i$ is the result of the multiplication of logits with softmax function. We then normalized entropy values between 0 and 1 to obtain uncertainty values.

\subsection{Uncertainty-based ensemble learning}
We evaluated four variants of uncertainty-based ensemble learning described below.

\noindent{\textbf{Uncertainty lowest (ul)}}.
In this variant, we select the model with the lowest uncertainty value and infer the label based on that value.

\noindent{\textbf{Uncertainty threshold (ut)}}. This variant is similar to \textbf{ul} by choosing the lowest value but below the threshold. For the uncertainty values above the threshold, labels are inferred based on the mean ensemble. The threshold is searched using a grid search from 0.11 to 0.9 with a step size of 0.01.

\noindent{\textbf{Uncertainty weighted (uw)}}. We calculated the inverse of uncertainty ($1/$uncertainty) as the weights, normalized the weights per model, then multiplied each class model probability with their normalized weights and used the maximum one to infer the label.

\noindent{\textbf{Confidence weighted (cw)}}. We calculated the confidence score as a complement to 1 of uncertainty ($1-$uncertainty) and then normalized them for all samples per model. Similar to \textbf{uw}, we multiply each class model probability with their normalized weights and use the maximum one to infer the label.

\section{Experiments}
\label{sec:pagestyle}
\subsection{Tasks and Datasets}
Five tasks and ten datasets are evaluated using speaker-independent criteria except for SER-EMNS and SR-RAVDESS. EMNS uses a single speaker while in SR-RAVDESS the goal is to predict the speaker ID.

\noindent{\textbf{Speech emotion recognition (SER)}} is a task to classify speech signals into a category of emotions. For this SER tasks, we evaluated IEMOCAP \cite{Busso2008}, EMNS \cite{Noriy2023}, TurEV \cite{Canpolat2020}, KBES \cite{Billah2023}, Polish \cite{mmiesikowska2020}, and TTH \cite{Thi2023} datasets. The choices of these datasets are based on the purpose of the ensemble learning to improve performance scores. We choose datasets with unweighted accuracies around 70\% or lower from the previous study \footnote[1]{https://github.com/felixbur/nkululeko/tree/main/data}. We evaluated four emotion categories except for the Polish dataset, which contains only three categories (`anger,' `neutral,' and `fear'). The four emotions are `neutral,' `happiness,' `anger,' and `sadness' for IEMOCAP, EMNS, KBES, and TTH, and `angry,' `calm,' `happy,' and `sad' for TurEV. EMNS dataset is balanced using SMOTE (\cite{Lemaitre2017}) during the training process, while others are kept in the original forms.

\noindent{\textbf{Non-verbal emotion recognition (NVER)}} is a task similar to SER, but instead of verbal speech signals, it contains non-verbal voice signals such as crying, laughing, screaming, or other non-verbal phrases. We evaluated two datasets for the NVER task, VIVAE \cite{Holz2022} and JNV \cite{Xin2024}. We evaluated four emotion categories, i.e., `anger,' `fear,' `pleasure,' and `surprise,' for VIVAE and `angry,' `disgust,' `surprise,' and `sad,' for JNV.

\noindent{\textbf{Speaker recognition (SR)}} determines who is speaking in an audio signal. We evaluated RAVDESS \cite{Livingstone2018}, which is a dataset of 24 actors intended to perform ten different emotions.

\noindent{\textbf{Gender prediction (GP)}} is a task to predict the gender of a speaker, male or female. We evaluated the RAVDESS (12 male and 12 female) dataset \cite{Livingstone2018}, which is already annotated with gender information.

\noindent{\textbf{Laughter classification (LC)}} is a task to classify laughter into different categories, happy and evil laughter data, distinguishing ``laugh with" from ``being laughed at." We evaluated the EvilLaughter database \cite{Dusterhoft2023}, sampled from Audioset \cite{Gemmeke2016} with examples of happy and evil laughter occurrences. The dataset contains 90 samples, each 45 per laughter category.

\subsection{Acoustic Features}
We evaluated three acoustic features (audmodel, hubert, and wavlm) for the SER and NVER experiments and two acoustic features (os, praat) for the SR, GP, and LC experiments. Description of the acoustic features are as follows:

\noindent{\textbf{audmodel}} \cite{Wagner2022a} is a finetuned wav2vec2-large-robust model on the MSP-PODCAST dataset \cite{Lotfian2019}. The feature is known to be useful for SER tasks \cite{Atmaja2023c}, as well as vocal bursts emotion recognition \cite{Atmaja2023}. We used audmodel as feature extractor to extract 1024-dimensional feature vector for each speech signal.

\noindent{\textbf{hubert}} \cite{Hsu2021} is self-supervised speech representation
learning by masked prediction of hidden units. Specifically, we used the \textit{hubert-large-ll60k} version (without finetuning) to extract a 1024-dimensional feature vector for each speech signal. This feature type achieved the highest SER score in the 2021 SUPERB benchmark \cite{Yang2021}.

\noindent{\textbf{wavlm}} is a large-scale self-supervised pre-training for full stack speech processing \cite{Chen2021}. We evaluated the large version of WavLM for SER and NVER tasks. This feature has also achieved top performance on major SER datasets \cite{Atmaja2022h}.

\noindent{\textbf{os}} (openSMILE) is the Munich versatile and fast open-source audio feature extractor \cite{Eyben2015b}. Specifically, we used the \text{eGeMAPSv0} version \cite{Eyben} to extract an 88-dimensional feature vector for each speech signal.

\noindent{\textbf{praat}} is a software package for doing phonetic analysis of speech \cite{Boersma2001}. We used Praat via parselmouth-praat \cite{Jadoul2018} to extract pitch, the standard deviation of pitch, hnr, jitter, shimmer, formants, and other features of each speech signal. The feature dimension is 39.

All features are scaled using a standard scaler before being used for classification.

\subsection{Classifiers}
A Support Vector Machine (SVM) for classification has been used for all tasks with regularization is set to 1.0 for all tasks.

\subsection{Evaluation Metrics}
All tasks are evaluated with unweighted accuracy (UA) and weighted accuracy (WA) scores. UA equals to unweighted average recall (UAR) or balanced accuracy while WA equals to overall accuracy. In the case of class-balanced distribution (of target class), like in SR and GP tasks, score of UA and WA are the same.

We implemented dataset processing, acoustic feature extraction, classifier, and evaluation metrics using the Nkululeko toolkit \cite{Burkhardt2022}. Samples of configuration files to perform the experiments in this study are openly accessible \footnote[2]{\url{https://github.com/bagustris/nkululeko_ensemble_speech_classficcation}}.

\section{Results and Discussion}
\label{sec:result}
\subsection{Visualizing Effect of Uncertainty}
\begin{figure}[htbp]
    \centering
    \includegraphics[width=0.5\textwidth]{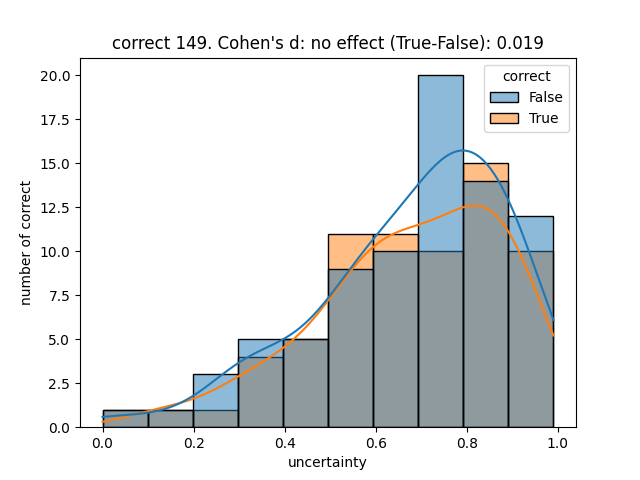}
    \includegraphics[width=0.5\textwidth]{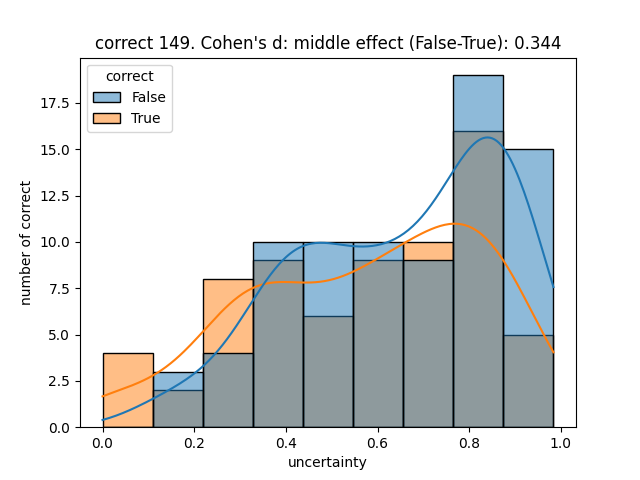}
    \includegraphics[width=0.5\textwidth]{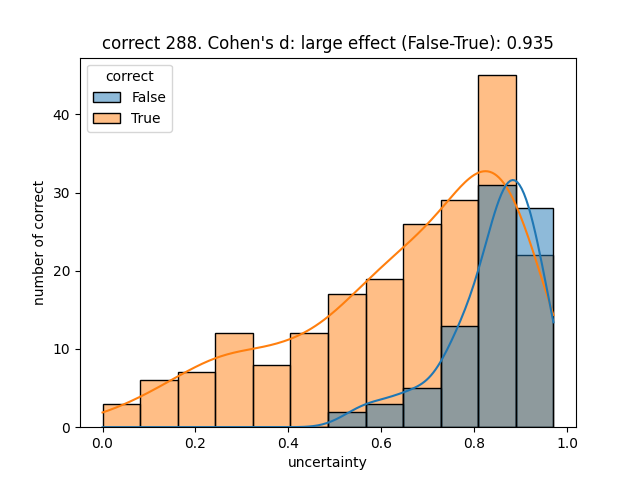}
    \caption{Samples Cohen's D effect with no effect (top, EMNS with os), middle effect (middle, KBES with wavlm), and large effect (bottom, SR-RAVDESS with praat)}
    \label{fig:cohen_effect}
\end{figure}

We can observe the effect of using uncertainty calculation on the correctness of predictions to gain insight into the usefulness of uncertainty quantification in speech classification. We found on the evaluated datasets the Cohen's D effect could be categorized into three: no effect, middle effect, and large effect (Fig. \ref{fig:cohen_effect}), with most data having a large effect. Given the large effect of Cohen's D, we expect that utilizing uncertainty calculation could be useful in ensemble learning.

The size of Cohen's D indicates the correlation between the correctness of predictions and the uncertainty values. The larger the effect size, the higher the correlation, meaning the uncertainty values are more reliable indicators of prediction correctness. Datasets with large Cohen's D effect show promising potential for using uncertainty-aware ensemble methods to improve overall classification performance.

%  explain correlation of Cohen -D with uncertainties
% the smaller means less useful to use uncertainty
% the larger the effect the more correlation between correctness and uncertainty

% \begin{figure}[htbp]
%   \centering
%   \includegraphics[width=0.5\textwidth]{figs/emns_aud_middle.png}
%   \includegraphics[width=0.5\textwidth]{figs/kbes_wavlm_middle.png}
%   \caption{Samples of Cohen's D effect with middle effect; top: bottom:}
%   \label{fig:middle_effect}
% \end{figure}

% explain Fig 3 top right or bottom-left: the uncertainty!
% if we only accept uncertainty less than 0.5 or 0.4 (bottom-left), 
% we would to get 100% accuracy, but the number of data will be small
% this is why we propose to use threshold

Fig. \ref{fig:cohen_effect} for the large effect (bottom) can be used to explain the concept of uncertainty. If we only accept uncertainty less than 0.5, we would get 100\% accuracy by choosing the lowest uncertainty score to infer the label, but the number of data will be small. This is why we propose to use a threshold. Based on our empirical study, the mean ensemble works better than other ensembles to infer labels for uncertainty values above the threshold.

% \begin{figure*}[htbp]
%   \centering
%   \includegraphics[width=0.5\textwidth]{figs/iemocap_aud.png}\includegraphics[width=0.5\textwidth]{figs/ravdess_speaker_praat.png}
%   \includegraphics[width=0.5\textwidth]{figs/polish_wavlm_large.png}\includegraphics[width=0.5\textwidth]{figs/tth_aud_large.png}
%   \caption{Samples of Cohen's D with large effect; top-left: IEMOCAP with AUDMODEL, top-right: RAVDESS (Speaker Recognition) with PRAAT; bottom-left: Polish dataset with WavLM, bottom-right: TTH with AUDMODEL }
%   \label{fig:large_effect}
% \end{figure*}

\subsection{Baseline: Single Models}
We measure the performance of single models to compare them with ensemble learning. The performance results of a single model are shown in Table \ref{tab:single_models}. The audmodel generally performs well across SER and NVER tasks. WavLM obtained the highest scores on SER-KBES and SER-Polish. Performance varies significantly across datasets, suggesting that some are more challenging than others. For NVER tasks, performance on the JNV dataset is notably higher than on the VIVAE dataset. The `os' approach generally shows higher performance than `praat' for the SR, GP, and LC tasks; for the GP task, both `os' and `praat' approaches achieve comparable high performance. The LC task shows lower performance compared to other tasks, with significant differences between UA and WA scores, perhaps due to the low amount of data.

% Creating the table
\begin{table}[htbp]
    \centering
    \caption{ Performance of single models; bold are the highest}
    \begin{tabular}{l c c c c  c c}
    \hline
 Task-dataset & \multicolumn{2}{c}{audmodel} & \multicolumn{2}{c}{hubert} & \multicolumn{2}{c}{wavlm} \\
    & UA & WA & UA & WA & UA & WA \\
    \hline
    \textbf{SER}-  & & & & & & \\
 IEMOCAP & \B 75.1 & \B 73.5 & 69.9 & 69.1   & 73.1 & 73.1 \\
 EMNS    & \B 53.7 & \B 58.3 & 48.3 & 51.0 & 40.1 & 46.3 \\
 TurEV   & \B 57.9 & \B 57.8 & 36.8 & 36.8 & 50.9 & 50.9 \\
 KBES    & 77.5 & 80.0 & 75.8 & 78.0 & \B 79.2 & \B 81.9 \\
 Polish  & 65.5 & 65.5 & 57.6 & 56.7 & \B 68.8 & \B 68.8 \\
 TTH     & \B 47.5 & \B 81.2 & 43.9 & 79.4 & 45.1 & 80.2 \\ 
    \hline
    \textbf{NVER}- & & & & & & \\
 VIVAE   & \B 63.6 & \B 62.9 & 57.5 & 56.9 & 59.6 & 58.9 \\
 JNV     & \B 81.7 & \B 79.4 & 74.1 & 70.5 & 74.2 & 67.6 \\
    \hline
    & \multicolumn{3}{c}{os} & \multicolumn{3}{c}{praat} \\
    & \multicolumn{3}{c}{UA~~~~WA} & \multicolumn{3}{c}{UA~~~WA} \\
    \hline
    \textbf{SR} & \multicolumn{3}{c}{\B 92.7~~~~\B 92.7} & \multicolumn{3}{c}{71.5~~~~71.5} \\
    \hline
    \textbf{GP} & \multicolumn{3}{c}{\B 98.9~~~~\B 98.9} & \multicolumn{3}{c}{97.2~~~~97.2} \\
    \hline
    \textbf{LC} & \multicolumn{3}{c}{{\B 50.0}~~~~21.4} & \multicolumn{3}{c}{43.9~~~~\B 50.0} \\
    \hline
    \end{tabular}
    \label{tab:single_models}
\end{table}

\subsection{Ensemble Learning Results}
Table \ref{tab:ensemble} presents results for ensemble learning, both uncertainty-based and conventional methods (mean, max). The uncertainty-based approaches (ul, ut, uw, cw) generally perform better than single models, with {\B ut} and {\B uw} performing the best on average across tasks. Conventional ensembles (mean, max) also improve over single models but are often outperformed by uncertainty-based approaches. Based on the occurrences of high scores compared to single models, {\B ut} and {\B uw} achieved the most improvements of combined UA an WA scores (17 higher scores, in bold), followed by {\B ul} (15 high scores), {\B cw} and mean (14), and max (12).

In many cases, the combination of all three modalities (aud+hub+wav) tends to perform better than individual modality pairs, suggesting that multimodal approaches can be beneficial for these tasks. However, in such cases as SER-EMNS, SER-TurEV, and SER-KBES, the use of two models for fusion leads to better results compared to three models fusion. This indicates a need to explore different model combinations for optimal ensemble learning.

For each task, different results are obtained. KBES dataset obtained the stronger results for SER. NVER tasks are evaluated on VIVAE and JNV datasets, with JNV showing higher accuracies overall. The SR-RAVDESS and GP-RAVDESS tasks show very high accuracies, with GP achieving perfect scores across all metrics for all ensemble methods. The LC-Laughter task shows lower accuracies compared to other tasks, indicating it might be a more challenging classification problem. 

\begin{table*}[htbp]
    \centering
    \caption{Performance of ensemble learning; ul: uncertainty-lowest, ut:uncertainty-threshold, uw: uncertainty-weighted, cw: confidence-weighted; aud:audmodel, hub:hubert, wav:wavlm; bold: higher than the best single model}
    \label{tab:ensemble}
    \begin{tabular}{l c c c c c c c c c c c c c}
    \hline
 Task-dataset & \multicolumn{2}{c}{ul} & \multicolumn{2}{c}{ut} & \multicolumn{2}{c}{uw} & \multicolumn{2}{c}{cw} & \multicolumn{2}{c}{mean} & \multicolumn{2}{c}{max}\\
    & UA & WA & UA & WA & UA & WA & UA & WA & UA & WA & UA & WA \\
    \hline
    \hline
 SER-IEMOCAP & & & & & & & & & & & & \\
 aud+hub     & \B 75.3   & \B 75.8 & \B 75.5 & \B 74.4 & \B 75.4 & \B 75.2   & \B 75.5 & \B 74.3     & \B 75.4 & \B 74.3 & \B 75.2 & \B 74.0 \\
 hub+wav     & 72.3  & 71.8 & 72.4 & \B 74.4 & 72.1 & 71.8   & 71.9 & 71.6       & 72.1 & 71.8   & 72.0 & 71.6 \\
 aud+hub+wav & \B 76.2   & \B 74.9 & \B 76.2 & \B 75.2 & \B 75.9 & \B 75.1   & \B 76.1 & \B 75.2     & \B 75.4 & \B 74.8 & \B 75.7 & \B 74.7 \\
 SER-EMNS & & & & & & & & & & & & \\
 aud+hub     & 51.1 & 56.4  & 50.4 & 55.7 & 50.4 & 55.7 & 51.2 & 56.4 & 50.4 & 55.7 & 52.0 & 57.0 \\
 hub+wav     & 53.2 & \B 58.4   & 42.7 & 48.3 & \B 57.0 & \B 62.4   & \B 55.8 & \B 61.1 & \B 57.4 & \B 62.4 & \B 55.8 & \B 56.1 \\
 aud+hub+wav & 49.4 & 55.0  & 50.6 & 56.4 & 50.6 & 56.4 & 50.2 & 55.7 & 49.8 & 55.0 & 52.4 & 57.7 \\
 SER-TurEV & & & & & & & & & & & & \\
 aud+hub     & 57.3  & 57.3 & \B 58.8 & \B 58.8 & \B 58.2 & \B 58.2 & 57.9 & 57.9 & \B 58.2 & \B 58.2 & 57.6 & 57.6 \\
 hub+wav     & 47.6  & 47.6 & 47.6 & 47.6 & 47.0 & 47.0 & 46.6 & 46.6 & 46.3 & 46.3 & 47.3 & 47.3 \\
 aud+hub+wav & 57.3  & 57.3 & 57.9 & 57.9 & 57.6 & 57.6 & 57.6 & 57.6 & 56.1 & 56.1 & 57.3 & 57.3 \\
 SER-KBES & & & & & & & & & & & & \\
 aud+hub     & 78.3  & 81.9 & 79.2 & \B 82.9 & 79.2 & \B 82.9    & 79.2 & \B 82.9    & 79.2 & \B 82.9    & 78.3 & 81.9 \\
 hub+wav     & 79.2  & 81.8 & 79.2 & 81.9 & 79.2 & 81.9  & 79.2 & 81.9  & 79.2 & 81.9   & 79.2 & 81.9 \\
 aud+hub+wav & 78.3  & 81.9 & 78.3 & 81.9 & 78.3 & 81.9  & 78.3 & 81.9   & 77.5 & 81.0   & 78.3 & 81.9 \\
 SER-Polish & & & & & & & & & & & & \\
 aud+hub     & 65.6  & 65.6  & 67.8 & 67.8 & 67.8 & 67.8 & 67.8 & 67.8 & 67.8 & 67.8 & 67.8 & 67.8 \\
 hub+wav     & 67.8  & 67.8  & 67.8 & 67.8 & 66.7 & 66.7 & 67.8 & 67.8 & 67.8 & 67.8 & 67.8 & 67.8 \\
 aud+hub+wav & 67.8  & 67.8  & \B 68.9 & \B 68.9 & \B 68.9 & \B 68.9 & 67.8 & 67.8 & 67.8 & 67.8 & 67.8 & 67.8 \\
 SER-TTH & & & & & & & & & & & & \\
 aud+hub     & 45.6  & 80.5 & 45.6 & 80.5 & 45.6 & 80.6  & 45.6 & 80.6 & 45.6 & 80.7 & 45.5 & 80.6 \\
 hub+wav     & 44.4  & 79.8 & 44.6 & 79.9 & 44.1 & 79.9  & 44.1 & 79.9 & 44.1 & 79.9 & 44.1 & 79.9 \\
 aud+hub+wav & 45.6  & 80.6 & 45.6 & 80.6 & 45.1 & 80.4  & 45.1 & 80.5 & 44.9 & 80.4 & 45.4 & 80.5 \\
    \hline
 NVER-VIVAE & & & & & & & & & & & & \\
 aud+hub     & \B 67.9 & \B 67.5 & \B68.6 & \B 68.2 & \B 68.6 & \B 68.2 & \B 67.9 & \B 67.5  & \B 68.6 & \B 68.2 & \B 67.3 & \B 66.9 \\
 hub+wav     & \B 67.9 & \B 67.5 & \B64.9 & \B 64.2 & \B 64.9 & \B 64.2 & 63.1 & 62.3  & \B 64.9 & \B 64.2 & 63.1 & 62.3 \\
 aud+hub+wav & \B 68.2 & \B 67.5 & \B68.8 & \B 68.2 & \B 68.2 & \B 67.5 & \B 69.5 & \B 68.9  & \B 68.2 & \B 67.5 & \B 68.1 & \B 67.5 \\
 NVER-JNV & & & & & & & & & & & & \\
 aud+hub     & \B 84.4   & \B 85.3 & \B 84.4 & \B 85.3 & \B 83.1 & \B 82.4   & \B 83.1 & \B 82.4 & \B 83.1 & \B 82.4 & \B 83.1 & \B 82.4 \\
 hub+wav     & 75.6  & 70.6 & 79.5 & 79.4 & 75.6 & 70.6  & 75.6 & 70.6 & 75.6 & 70.6 & 75.6 & 70.6 \\
 aud+hub+wav & 80.8  & \B 82.4 & \B 81.8 & 79.4 & \B 81.8 & 79.4 & \B 83.1 & \B 82.4 & 78.2 & 76.5   & 79.5 & 79.4 \\
    \hline
 SR-RAVDESS & & & & & & & & & & & & \\
 os+praat & 92.7 & 92.7 & \B 94.4 & \B 94.4 & \B 94.1 & \B 94.1 & \B 94.1 & \B 94.1 & \B 94.4 & \B 94.4 & \B 93.1 & \B 93.1 \\
    \hline
 GP-RAVDESS & & & & & & & & & & & & \\
 os+praat & \B 100 & \B 100 & \B 100 & \B 100 & \B 100 & \B 100 & \B 100 & \B 100 & \B 100 & \B 100 & \B 100 & \B 100 \\
    \hline
 LC-Laughter & & & & & & & & & & & & \\
 os+praat    & \B 54.5 & 28.6 & \B 59.1 & 35.7 & 25.8 & 21.4 & 42.4 & 28.6 & 34.8 & 35.7 & 34.8 & 35.7 \\
    \hline
    \end{tabular}
\end{table*}

% most data show large cohen-d effect, including TTH dataset, but not all data shows improvement on using uncertainty-based fusions.
Finally, although most datasets show large Cohen's D effects when comparing the distribution of correctness from uncertainty calculation, not all datasets show improvement when using uncertainty-based fusion methods, e.g., the TTH dataset. The calculation of UQ, which is based on entropy, can be extended to other methods, e.g., Monte Carlo dropout, to further observe uncertainty estimation and fusion performance. The ensemble learning could also be extended to regression tasks in addition to speech classification.

\section{Conclusions}
\label{sec:conclusions}
In this paper, we investigated uncertainty-based ensemble learning for speech classification. We evaluated four variants of uncertainty-based ensemble learning methods and compared them with single models and conventional ensemble learning methods. The experimental results show that the uncertainty threshold ({\B ut}) and uncertainty weighted ({\B uw}) methods obtained higher gain than other methods in terms of the quantity of improved UA/WA scores. The threshold in {\B ut}, however, is searched manually for each dataset, which is impractical for real applications. The uncertainty-weighted method, on the other hand, did not require manual threshold selection and achieves similar performance to the {\B ut} method. 

% \section{Acknowledgment}
% \label{sec:ack}
% \pagebreak
% This paper is partly based on results obtained from a project, JPNP20006, commissioned by the New Energy and Industrial Technology Development Organization (NEDO), Japan.

% -------------------------------------------------------------------------
\bibliographystyle{IEEEbib}
\small
\bibliography{ensemble}

\end{document}